\documentclass[a4paper,fleqn,usenatbib]{mnras}

\usepackage{txfonts}

\usepackage[T1]{fontenc}
\usepackage{ae,aecompl}
\usepackage{longtable}
\usepackage{color}
\usepackage{array}
\usepackage{graphicx}
\usepackage{caption}
\usepackage{url}
\usepackage{aecompl}
\usepackage{subfig}
\usepackage{float}

\newcommand{\RNum}[1]{\uppercase\expandafter{\romannumeral #1\relax}}

\title[Outer Envelopes of Globular Clusters]{The Outer Envelopes of Globular Clusters. \RNum{1}. NGC 7089 (M2)\thanks{This paper includes data gathered with the 6.5 meter Magellan Telescopes located at Las Campanas Observatory, Chile.}}
\author[P. B. Kuzma et al.] {P. B.~Kuzma,$^{1}$\thanks{E-mail: pete.kuzma@anu.edu.au}, G. S.~Da~Costa$^{1}$, A. D.~Mackey$^{1}$, T. A.~Roderick$^{1}$\\
$^{1}$Research School of Astronomy and Astrophysics, Australian National University, Canberra, ACT 2611, Australia
}

\date{Accepted XXX. Received YYY; in original form ZZZ}

\pubyear{2016}

\begin{document}
\label{firstpage}
\pagerange{\pageref{firstpage}--\pageref{lastpage}} 
\maketitle

\begin{abstract}
We present the results of a wide-field imaging survey of the periphery of the Milky Way globular cluster NGC 7089 (M2). Data were obtained with MegaCam on the Magellan Clay Telescope, and the Dark Energy Camera on the Blanco Telescope. We find that M2 is embedded in a diffuse stellar envelope extending to a radial distance of at least $\sim 60\arcmin$ ($\sim 210$ pc) -- five times the nominal tidal radius of the cluster. The envelope appears nearly circular in shape, has a radial density decline well described by a power law of index $\gamma = -2.2 \pm 0.2$, and contains approximately $1.6\%$ of the luminosity of the entire system. While the origin of the envelope cannot be robustly identified using the presently available data, the fact that M2 also hosts stellar populations exhibiting a broad dispersion in the abundances of both iron and a variety of neutron capture elements suggests that this object might plausibly constitute the stripped nucleus of a dwarf Galaxy that was long ago accreted and destroyed by the Milky Way.
\end{abstract}


\section{Introduction}
In the $\Lambda$CDM cosmological model, present-day large galaxies form hierarchically \citep[e.g.,][]{2002NewA....7..155S}. Dark matter clumps merge and combine at early times to form protogalaxies, which themselves merge into larger systems, and so on. Stellar halos around large galaxies are thought to arise as a by-product of these processes \citep[e.g.,][]{2005ApJ...635..931B,2010MNRAS.406..744C}; the growth of this component continues even at late times via the accretion of dwarf galaxies into massive systems, contributing stars and globular clusters into the diffuse halo region. The seminal work of \cite{1978ApJ...225..357S} provided some of the first observational evidence for this scenario in the Milky Way, by demonstrating that globular clusters outside the solar circle do not exhibit the correlation between Galactocentric distance and metallicity observed among innermost globular clusters. More recent work has revealed that a substantial fraction of Milky Way globular clusters follow a clear age-metallicity relationship that is consistent with their formation in external systems \citep[e.g.,][]{2009ApJ...694.1498M,2010ApJ...708..698D, 2013MNRAS.436..122L}; there are also distinct similarities between many outer halo globular clusters in the Milky Way and globular clusters seen in nearby dwarf galaxies \citep[e.g.,][]{2004MNRAS.355..504M,2005MNRAS.360..631M}. Collectively this evidence suggests that the current halo globular cluster population is a mixture of objects of extra-Galactic origin and those that formed in the Milky Way.

Direct evidence for the build up of the Galactic halo via the accretion of smaller galaxies came with the serendipitous discovery of the disrupting Sagittarius dwarf \citep{1994Natur.370..194I}. The stream associated with this system can be traced in a complete loop around the Milky Way \citep[e.g.,][and references therein]{2009ApJ...700.1282Y}, and a number of globular clusters have been linked with the dwarf -- either directly \citep[namely M54, Arp 2, Terzan 7 and Terzan 8,][]{1995AJ....109.2533D}, or through possible association with the stream \citep{1995MNRAS.277..781I,2003AJ....125..188B,2004ApJ...601..242M,2010ApJ...718.1128L}. 

However, despite the interaction between Sagittarius and the Milky Way unravelling before us, and the discovery, to date, of nearly two dozen much smaller stellar streams, there is an apparent dearth of large-scale substructures in the Milky Way halo when compared to the situation observed in our neighbouring spiral galaxy, M31. The Pan-Andromeda Archaeological Survey \citep[PAndAS;][]{2009Natur.461...66M} has utilised deep wide-field imaging to reveal that the M31 halo contains an abundance of large streams and over-densities \citep[e.g.,][]{2014ApJ...780..128I}, as well as a substantial globular cluster population extending to very large Galactocentric radii \citep{2014MNRAS.442.2165H}. Many of these remote globular clusters are spatially coincident with, and share the same velocity as, underlying stellar streams \citep{2010ApJ...717L..11M,2013MNRAS.429..281M,2014MNRAS.445L..89M,2013ApJ...768L..33V,2014MNRAS.442.2929V}, indicating that they were formed in satellite dwarfs that were subsequently accreted into the M31 halo. 

It is not clear whether the apparent lack of large streams in the Milky Way halo compared to the M31 halo reflects an intrinsic difference between the two galaxies, or is the result of observational bias.  Finding large-scale structures in M31 is certainly a considerably easier task than for the Milky Way -- the angle subtended by the M31 halo is tiny compared to the all-sky surveys required, at similar photometric depth, to probe to commensurate radii in the Milky Way halo. At present our best efforts come from major surveys such as the Sloan Digital Sky Survey \citep[SDSS,][]{1996AJ....111.1748F,1998AJ....116.3040G,2000AJ....120.1579Y} and the Pan-STARRS1 $3\pi$ survey \citep{2012ApJ...750...99T}; however these are comparatively shallow and only trace the Milky Way halo out to $\sim 30$ kpc {\it at high contrast}. Probing to larger distances requires the use of rare tracers, such as blue horizontal branch stars, RR Lyrae variables, or M giants, that are not necessarily well suited to detecting very low surface brightness substructures. 

One alternative possibility is to employ a deep targeted survey. Since most of the globular clusters in the outer M31 halo reside in or near stellar streams, there are globular clusters known to be embedded in the Sagittarius stream, and many other remote Milky Way clusters are hypothesised to be accreted objects, it is plausible that globular clusters in the Milky Way might act as efficient tracers for distant large scale halo structures. Indeed, an attempt to search for streams around a variety of Galactic globular clusters has been performed recently by \citet{2014MNRAS.445.2971C}. While between six and ten clusters in their sample of $23$ show promising evidence for minor stellar populations beyond their tidal radii, ultimately the lack of a sufficiently large field of view left the authors unable to draw any firm conclusions as to whether these populations might represent large streams, globular cluster tidal tails, or some other kind of extended structure. A handful of other similar studies have been performed in the past decade \citep[e.g.,][]{2000A&A...359..907L,2010AJ....139..606C,2010A&A...522A..71J}, and while some globular clusters have been reported to have tidal tails, no large scale streams have been discovered.

\begin{figure*}
  \begin{center}  
    \includegraphics{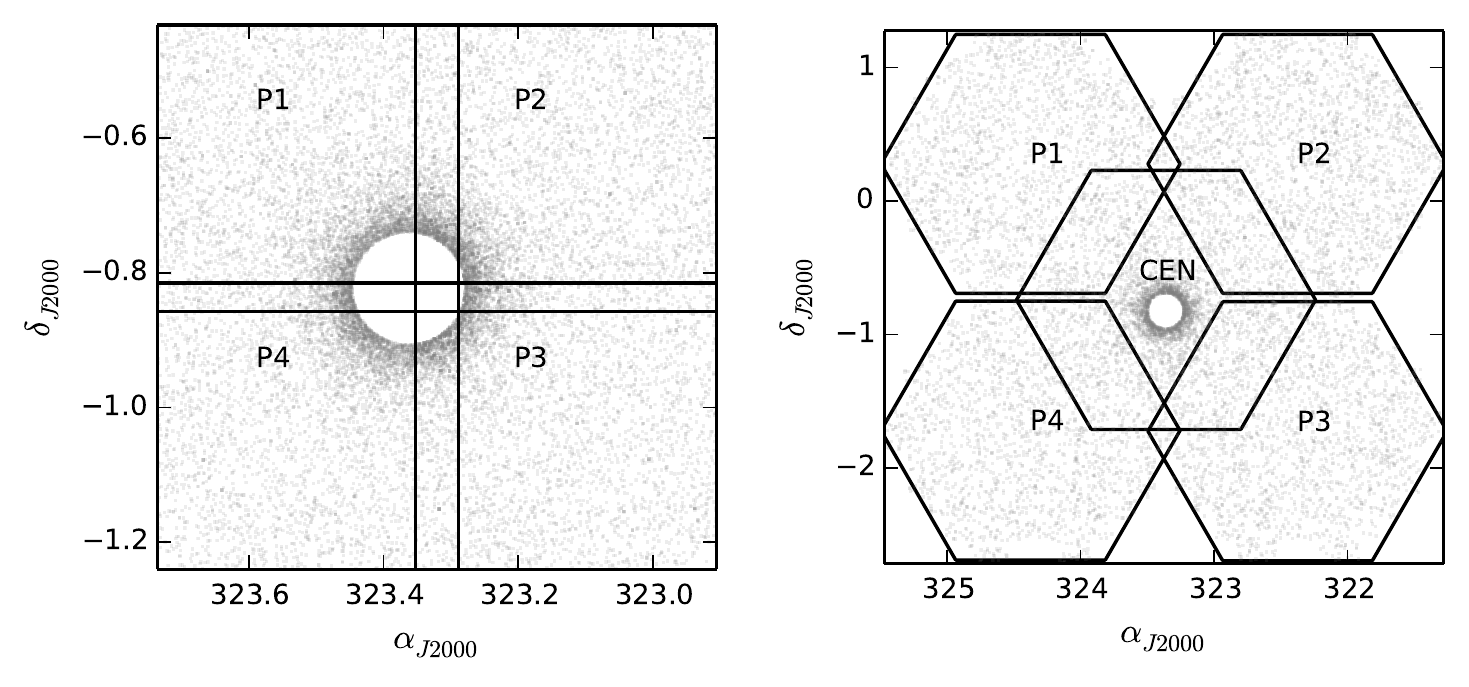}
  \end{center}
\caption{Our observed fields around M2 from MegaCam (left) and DECam (right). Detected sources are marked with grey points. The crowded central regions of the cluster have been excluded; the radii of the excluded regions are $5\arcmin$ for MegaCam and $7\arcmin$ for DECam.}
\label{fig:M2field}
\end{figure*}
 
We are conducting our own search for stellar streams in the outer Galactic halo by studying globular clusters and their surroundings. Modern wide-field mosaic imagers such as the Dark Energy Camera (DECam; \citealt{2015AJ....150..150F}) on the 4-m Blanco Telescope at Cerro Tololo Inter-American Observatory (CTIO) and MegaCam \citep{2015PASP..127..366M} on the 6.5-m Clay Telescope at Las Campanas Observatory (LCO), are perfect instruments for this task. We have predominantly targeted clusters that have properties indicative of a possible extra-Galactic origin. As well as large-scale streams belonging to destroyed dwarf galaxies, it is possible that we may reveal tidal tails that belong to the globular clusters themselves. Such structures are already known for several Galactic globular clusters -- the prototypes being Palomar 5 \citep[e.g.,][]{2001ApJ...548L.165O,2006ApJ...641L..37G,2009AJ....137.3378O} and NGC 5466 \citep[e.g.,][]{2006ApJ...637L..29B,2006ApJ...639L..17G}. They exhibit a characteristic two-arm structure, and have a width that is approximately that of the progenitor cluster. This differentiates them from debris due to a lost dwarf galaxy host, which is expected to be much broader on the sky such that it surrounds a cluster in all directions.

In this paper we report results for the first target of our survey, NGC 7089 (M2). This cluster possesses a variety of unusual characteristics, some of which are suggestive of an extra-Galactic origin. \citet{1995AJ....109.2553G} explored the outskirts of M2 through star counts from photographic plates and found indications of extended structure surrounding the cluster, including significant deviations in the radial density profile from the expected \citet{1962AJ.....67..471K} shape. They concluded it was likely that M2 possesses tidal tails. More recently, it has been revealed that M2 hosts stellar populations with a broad dispersion in iron abundance -- \citet{2014MNRAS.441.3396Y} detected a dominant peak at $[$Fe$/$H$] \approx -1.7$ and weaker peaks in the distribution at $[$Fe$/$H$] \approx -1.5$ and $-1.0$, though these results have been challenged by \citet{2016MNRAS.457...51L}. Furthermore, \citet{2014MNRAS.441.3396Y} also presented evidence for significant star-to-star variation in a number of neutron capture elements \citep[see also][]{2013MNRAS.433.1941L}, and variations in light element abundances have been found by \cite{2012A&A...548A.107L}. These properties are unusual, observed in only a handful of Galactic globular clusters. They have been reinforced photometrically -- precision multi-band measurements from the {\it Hubble Space Telescope} have revealed a complex colour-magnitude diagram that, in particular, exhibits multiple sub-giant branches \citep{2015MNRAS.447..927M}, corresponding well with the peaks in the metallicity distribution published by \citet{2014MNRAS.441.3396Y}. Combined, these properties render M2 rather similar to other anomalous massive clusters such as $\omega$ Cen and M54. The former has long been suggested as the remaining core of a long-defunct dwarf galaxy \citep[e.g.,][]{1993ASPC...48..608F}, while the latter resides at the centre of the Sagittarius dwarf {\citep[e.g.,][]{1995MNRAS.277..781I, 2000AJ....119.1760L}.

\section{Observations and Data reduction}
\subsection{Observations}
This work utilises two sets of observations, as summarised in Table \ref{tab:observations}. The first set was obtained with the MegaCam instrument on the 6.5m Magellan Clay telescope at LCO on 2013 September 10. MegaCam is a mosaic wide field imager that utilises 36 2048 x 4608 CCDs arranged in a $9 \times 4$ array, allowing for a $25\arcmin \times 25\arcmin$ field of view \citep{2015PASP..127..366M}. The binned pixel scale is $0.16$ arcsec/pixel. We obtained a mosaic of four pointings, with the cluster located in the corner of each field (Figure \ref{fig:M2field}; left) in order to maximise the area imaged around its outskirts. Each field was observed in the $\textit{g}$ and $\textit{i}$ bands for $3\times 90$ seconds and $3\times 300$ seconds respectively. The exposures were dithered to allow complete coverage by filling the gaps between the CCDs. Altogether, our four pointings cover a $0.8\degr \times 0.8\degr$ region centred on M2. The image quality during this set of observations varied, with that in $\textit{g}$ ranging between $0.6\arcsec - 0.9\arcsec$, and that in $\textit{i}$ between $0.5\arcsec - 0.9\arcsec$. Basic processing of the data -- bias subtraction, flat-fielding, astrometric calibration and image stacking -- was conducted using the MegaCam reduction pipeline\footnote{\url{http://hopper.si.edu/wiki/piper/Megacam+Data+Reduction}} available at the Harvard-Smithsonian Center for Astrophysics \citep[see][]{2015PASP..127..366M}. 

The second set of observations was obtained with DECam on the 4-m Blanco telescope at CTIO on 2013 September 26 as part of program number 2013B-0617 (PI: Mackey). DECam \citep{2015AJ....150..150F} is a mosaic wide field imager boasting a three square degree field, comprised of a roughly hexagonal arrangement of 62 2048 x 4096 CCDs with an associated pixel scale of $0.27$ arcsec/pixel. We observed five fields with DECam, arranged symmetrically in a cross-shape around M2 which was placed at the middle of the central field (see Figure \ref{fig:M2field}; right). Combined, our DECam data spans an approximately $13$ square degree region around M2. As with MegaCam, individual exposures at each pointing were dithered three times; each single exposure had an integration time of  $300$ seconds in both $\textit{g}$ and $\textit{i}$. For four of the five fields (CEN though to P3), the image quality was relatively consistent for both $\textit{g}$ ($\approx 1.1\arcsec - 1.2\arcsec$) and $\textit{i}$ ($\approx 1.0\arcsec - 1.1\arcsec$); however for the fifth field (P4) the image quality was noticeably poorer, particularly in the $\textit{g}$ band (see. Table \ref{tab:observations}). Basic processing of our DECam observations was carried out via the community pipeline\footnote{\url{http://www.ctio.noao.edu/noao/content/dark-energy-camera-decam}} \citep{2014ASPC..485..379V}. 

We note that MegaCam and DECam are complementary to each other for the present study. MegaCam has a comparatively higher spatial resolution and can perform deeper imaging in given exposure time than DECam, while DECam has a significantly larger field of view. Thus observations with MegaCam are perfect for exploring the crowded central regions of clusters, while DECam is ideal for exploring the vast space surrounding the cluster. Unless stated otherwise, the following discussion of our photometry procedures and data analysis is similar for both the MegaCam and DECam observations.

\begin{table*}

\begin{minipage}{160mm}
\begin{center}
\caption{Listing of the observations employed in this work.}

\label{tab:observations}
\begin{tabular}{@{}ccccccccccc}

\hline \hline
Camera&Date&Field&\multicolumn{2}{c}{Field Centre}&$N_{exp}$&Exp. Time & Filter &\multicolumn{3}{c}{Seeing (arcsec)}\\
&&Name&R.A. (J2000)& Dec. (J2000)&&per frame (s)&&1&2&3\\
\hline
MegaCam&2013 Sept 10&P1&21:34:03&-00:38:42&3&90&g&0.83&0.86&0.70\\
&&&&&3&300&i&0.89&0.60&0.56\\
&&P2&21:32:31&-00:38:42&3&90&g&0.65&0.63&0.66\\
&&&&&3&300&i&0.51&0.65&0.60\\
&&P3&21:32:31&-01:01:42&3&90&g&0.64&0.64&0.69\\
&&&&&3&300&i&0.56&0.58&0.64\\
&&P4&21:34:03&-01:01:42&3&90&g&0.91&0.80&0.86\\
&&&&&3&300&i&0.68&0.73&0.67\\
DECam&2013 Sept 26&CEN&21:33:27&-00:44:31&3&300&g&1.17&1.13&1.17\\
&&&&&3&300&i&1.03&1.05&1.02\\
&&P1&21:29:30&00:16:38&3&300&g&1.16&1.11&1.19\\
&&&&&3&300&i&1.08&1.00&1.02\\
&&P2&21:37:31&00:16:51&3&300&g&1.10&1.04&1.05\\
&&&&&3&300&i&1.00&1.00&1.16\\
&&P3&21:29:30&-01:43:15&3&300&g&1.06&1.09&1.09\\
&&&&&3&300&i&1.09&1.00&1.17\\
&&P4&21:37:30&-01:43:12&3&300&g&1.33&1.33&1.49\\
&&&&&3&300&i&1.11&1.09&1.24\\
\hline

\end{tabular}
\end{center}

\end{minipage}
\end{table*}

\subsection{Photometry}\label{sec:Photom}
Photometric measurements were obtained using Source Extractor\footnote{\url{https://www.astromatic.net/software/sextractor}} \citep[SExtractor;][]{1996A&AS..117..393B}. SExtractor is a software package that detects and performs photometry on sources in images, providing a variety of customisable parameters for the extraction. For this work, we utilised the aperture photometry feature from SExtractor to conduct our measurements. SExtractor was run twice on each image; the first run implemented a high detection threshold ($25\sigma$ above the mean pixel value) to find the brightest point sources in the field. These are predominately stars, and the measured median full-width at half-maximum (FWHM; $\bar{F}$) was used to define two aperture sizes ($1 \times \bar{F}$ \& $2 \times \bar{F}$) for a deeper subsequent application of SExtractor. This deeper extraction employed a detection threshold of $1.5\sigma$\footnote{The sigma is a true (local) pixel-to-pixel standard deviation. However, SExtractor has a number of algorithms in place to help reduce the number of spurious detections that appear at this level.}, a level that allows detection of the faintest objects in the image while maintaining a minimal number of spurious detections. This methodology delivered a photometric catalogue for all individual frames, and the corresponding stacked frames, per pointing per filter. 

We initially intended to work with the stacked images at each pointing.  However, we found that variations in the seeing between each of the individual exposures in a stack led to irregular variations in the stellar PSF across the field of view, resulting in sub-optimal photometry. This was true for both the MegaCam and DECam observations. Therefore, we chose to work with SExtractor measurements from the individual exposures, cross-matching the resulting catalogues and averaging the photometry for each given detection. This alleviated all of the problems arising from the use of the stacked images. We explored the possibility of systematically variable seeing across individual images, and while we observed a slight difference in some cases, we found that application of fixed apertures for each single frame provided sufficient photometric stability for our purposes.

An inevitable outcome of the photometric pipeline discussed above is the extraction of non-stellar objects, together with poor-quality and/or spurious detections. These are due to a variety of different source types -- background galaxies, blends between neighbouring stars, CCD defects, and, especially in single images, cosmic rays. It is imperative that we remove these unwanted detections from the SExtractor catalogues; to do this we employed a multi-step approach. First, SExtractor itself helped with the process via its internal diagnostic tools: the star/galaxy classifier and internal quality flags. The star/galaxy classifier is a number assigned to each detection that varies between zero and one, with zero referring to a definite galaxy and one to a definite star. Observing the distribution of this flag amongst fainter objects, there is a clear bimodal distribution at flag values at $0$ and $1$, as well as large number of stars gathering between a flag value of $0.5$ and $0.35$. With the possibility of losing photometric depth if performing too stringent a cut, we decided to remove objects with a star/galaxy classification of $0.35$ and below. However, none of the results presented here are strongly sensitive to the actual value adopted. The internal quality flags indicate the reliability of a given photometric measurement. A value of zero indicates a source that is located in a region bereft of nearby stars and that is not near the edge of a CCD. Lower quality photometry is represented by non-zero values\footnote{Please refer to the SExtractor manual for more information regarding the flags.}. We removed all objects with a non-zero quality flag.

\begin{figure}
  \begin{center}  
    \includegraphics{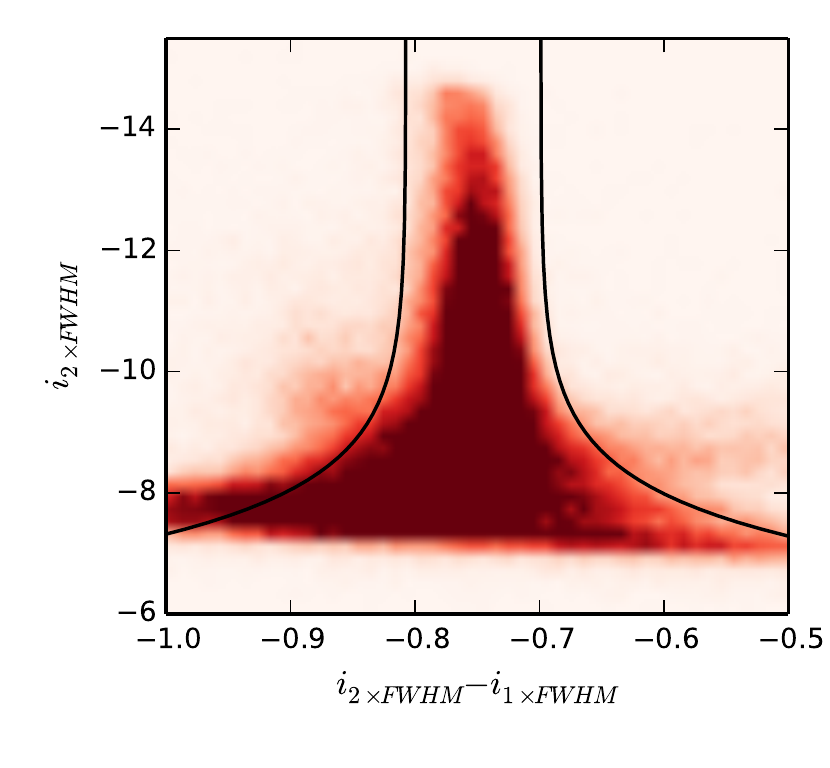}
  \end{center}
\caption{An example of the cleaning technique discussed in text. The boundaries delineate the regions where objects were removed from the sample as non-stellar detections -- most notably excluding the plume of galaxies on the left and probable cosmic rays on the right.}
\label{fig:clean}
\end{figure}

These combined SExtractor diagnostics were, however, insufficient to give satisfactory cleaning of galaxies and cosmic rays in the SExtractor output. We therefore implemented an additional step to help refine the stellar catalogues, by performing a cut based on the difference in magnitude between our two different aperture sizes. This difference in magnitudes should be consistent for stars (which share a similar light profile across a given image), but more negative for objects with a broader light profile (e.g., galaxies), and more positive for objects with a sharper light profile (e.g., cosmic rays).  In Figure \ref{fig:clean}, the brightest point sources populate a narrow magnitude difference and this spread becomes broader as the stars become fainter. This is indicative of uncertainties in the photometry increasing at lower magnitudes. Splitting the distribution in half about the median value of the aperture magnitude difference for stars $-13 \leq i \leq-11$, we fit an exponential curve to the right side of the distribution to define a boundary to eliminate non-stellar sources as well as point sources with unusually large uncertainties. This boundary was reflected about the median value to the left side of the distribution and objects that lay outside these boundaries (i.e., galaxies on the left and possible cosmic rays on the right side of the distribution) were removed. We performed this cut only on the $\textit{i}$-band catalogues, as the seeing was typically better in this filter than in the $\textit{g}$ filter. 

\begin{figure*}
  \begin{center}  
    \includegraphics{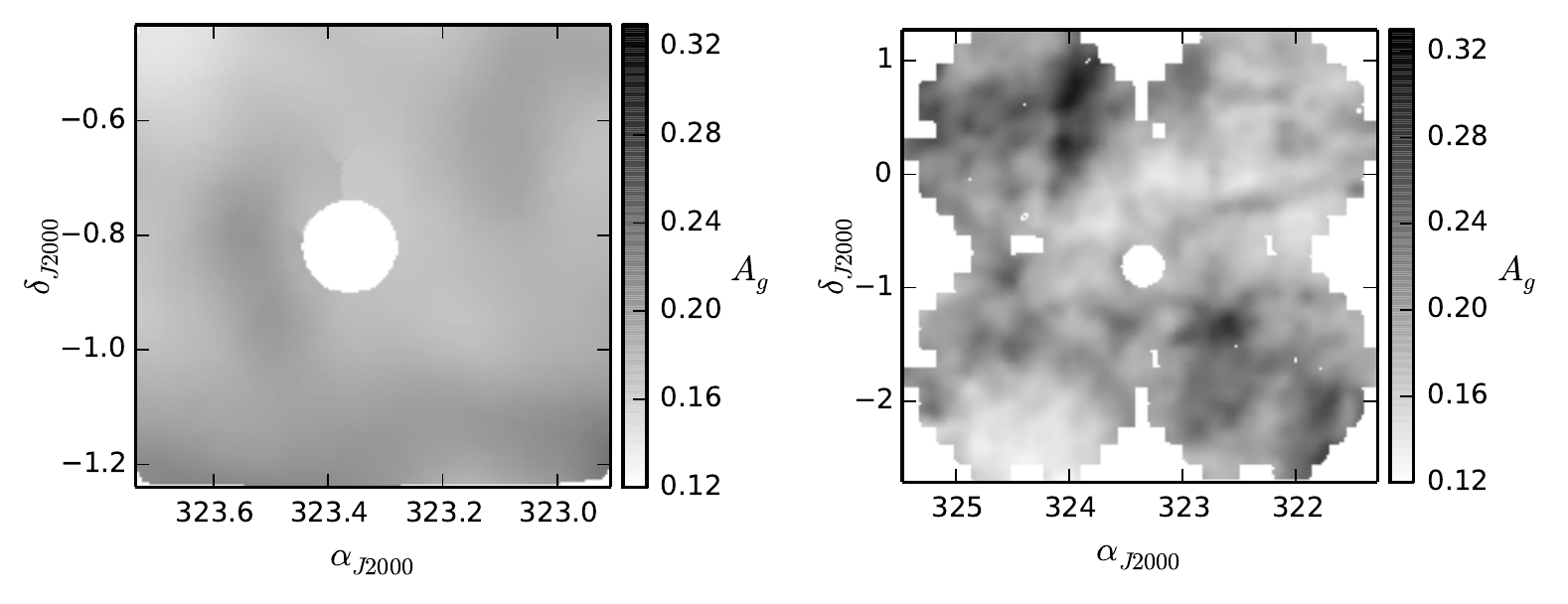}
  \end{center}
\caption{The $A_g$ extinction map for sources appearing in SDSS DR12 \citep{2014ApJS..211...17A}, based on the reddening maps of \citet{1998ApJ...500..525S}, and smoothed with a gaussian function of width $36\arcmin$. The MegaCam field of view is on the left, and DECam on the right. The excised inner cluster regions are the same as in Figure \ref{fig:M2field}. }
\label{fig:DUST}
\end{figure*}
 
We next desired to merge our individual catalogues and unify the photometric scales. First, for each $\textit{g}$ and $\textit{i}$ exposure pair (i.e., in a given field at matching dither points) the catalogues from our cleaning procedure were cross-matched using the command-line package Stilts \citep{2006ASPC..351..666T} to create lists containing stellar sources with good quality measurements and detections in both filters. Next, we cross-matched the three individual photometric catalogues for each pointing to create a single stellar catalogue for that field. For a given pointing the exposure with the deepest photometry was determined to be the master frame, and photometry from the remaining exposures was calibrated to the same scale as for the master. We did this by utilising the stars observed across multiple exposures to calculate the median photometric offsets, and then applied these to place all exposures on the same scale as the master. Once all the three exposures were on the same photometric scale, the catalogues were combined -- stars that were observed in either two or three of the images had their photometry calculated as the weighted mean of the SExtractor output photometry, using the inverse square of the uncertainties on the photometry reported by SExtractor as the weights. Finally, we repeated this process to merge all of the individual pointings for a given camera into a final catalogue, resulting in one catalogue for MegaCam observations and one for DECam observations. Overlapping regions between different fields were used to determine the offsets necessary to shift photometry for all pointings onto the same scale.

As a final step, we used photometry for the M2 region from the SDSS data release 12 \citep{2015ApJS..219...12A} to place our measurements onto an absolute scale. Stars recovered by our pipeline were cross-matched with the SDSS catalogue, and then used to fit to a linear relationship, plus a colour term, in order to transform from our instrumental magnitudes to the SDSS system. Table \ref{tab:SDSS} displays our zero points and coefficients for the colour term for both MegaCam and DECam. Once both the catalogues were calibrated to the SDSS photometry, we dereddened all magnitudes (denoted as $\textit{g}_{0}$ and $\textit{i}_{0}$) using the values contained in the SDSS catalogue, which originally come from the maps provided by \citet*{1998ApJ...500..525S}. Each star from our catalogues that was matched with an SDSS source was corrected by the corresponding reddening value listed in the SDSS catalogue, while those stars that did not have a match were given a correction that corresponded to that for the nearest star in the SDSS catalogue. Figure \ref{fig:DUST} shows the extinction across both fields of view for both cameras -- the reddening is mild but quite spatially variable.

\begin{table}
\begin{center}
\caption{The parameters used to calibrate our instrumental photometry to the SDSS system.}
\label{tab:SDSS}
\begin{tabular}{@{}cccc}
\hline \hline
Camera&Filter &\multicolumn{2}{c}{Calibration}\\
&&Zero Point & Colour Coeff.\\
\hline
MegaCam&g&$31.104\pm0.024$&$-0.050\pm0.009$\\
&i&$31.542\pm0.021$&$0.038\pm0.008$\\
DECam&g&$31.104\pm0.005$&$0.059\pm0.002$\\
&i&$31.165\pm0.003$&$0.072\pm0.001$\\
\hline
\end{tabular}
\end{center}
\end{table}

\subsection{Artificial Star Tests}
Since this study is concerned with searching for low surface brightness structures across large areas of sky, it was imperative to explore the completeness of our photometry as a function of magnitude and spatial position. If not properly accounted for, variable completeness levels across the different images in our mosaics could potentially result in detections of low surface brightness features that are not real. To quantify the completeness levels, we randomly placed 10000 artificial stars into each DECam field and 2000 artificial stars into each MegaCam, using the IRAF\footnote{\url{http://iraf.noao.edu/}} command \textit{mkobject}.The artificial stars had magnitudes between $17$ and $27.5$, with a higher proportion of stars at faint magnitudes to better reflect the luminosity function. After the artificial stars were placed in the fields, the images were run through the pipeline described in \S \ref{sec:Photom}, including the cleaning steps. The artificial stars were deemed as detected if they were found in the photometric catalogs after the cleaning steps. This process was repeated ten times per field, per camera, leading to 100000 simulated stars per field for DECam and 20000 stars for MegaCam.

The completeness function for each field, along with a corresponding fit using the interpolation model from \cite{1995AJ....109.1044F} is displayed in Figure \ref{fig:complete}.  To ensure uniformity across each of the two mosaics we decided to cut our catalogues at a level corresponding to $90\%$ completeness in the field with the shallowest photometry. With respect to our DECam measurements, we find the $\textit{g}-$band cut-off to be at $\textit{g} = 23.2$ and the $\textit{i}$-band cut-off to be at $\textit{i} = 22.3$. For our MegaCam measurements, the limits are at $\textit{g} = 23.6$ and $\textit{i} = 22.7$.

\begin{figure*}
  \begin{center}  
    \includegraphics{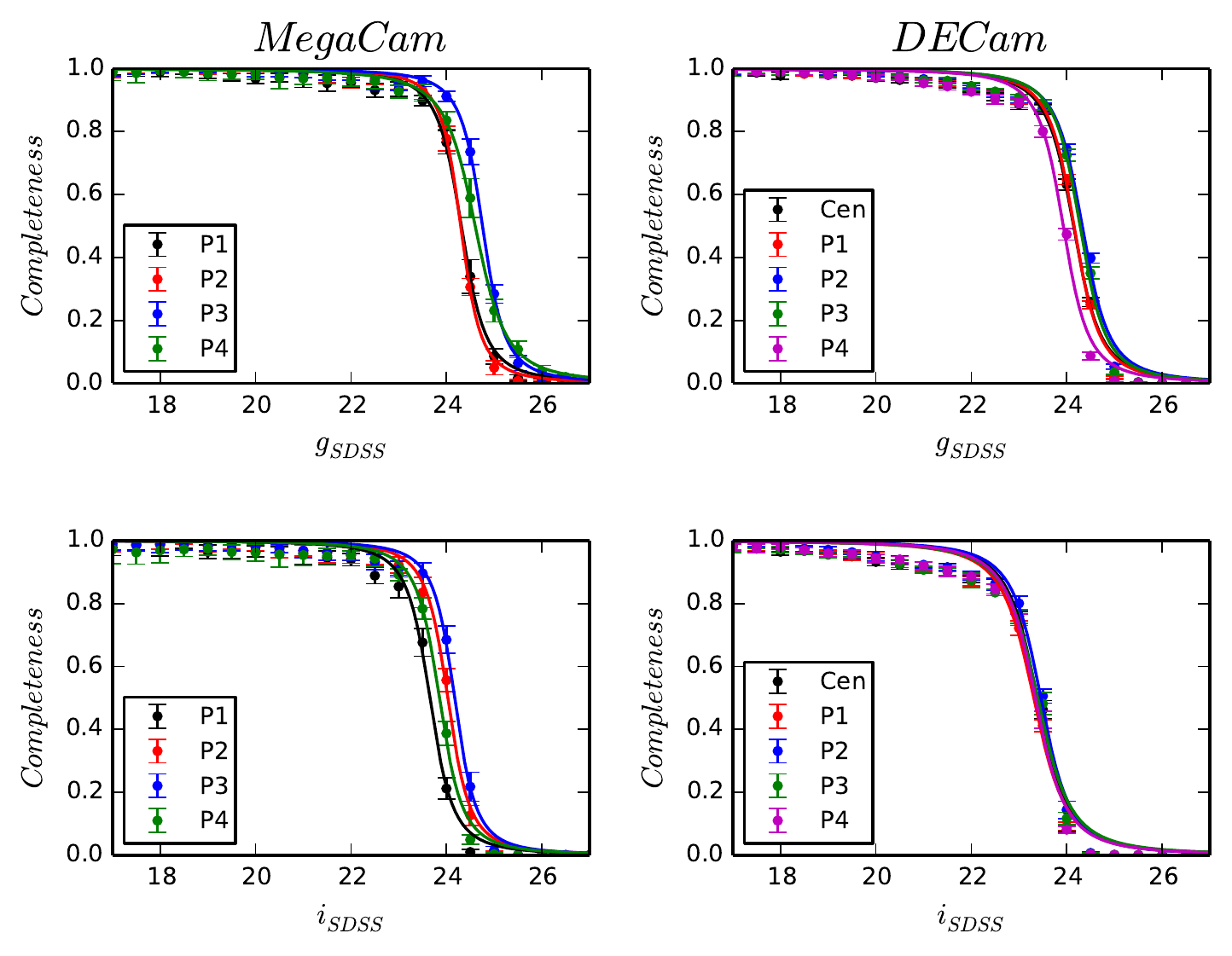}
  \end{center}
\caption{Completeness functions for each of our observed fields. The top row displays the $\textit{g}$-band observations and the bottom row presents the $\textit{i}$-band observations. The left column corresponds to MegaCam, and DECam is on the right. The completeness functions have been fit with the interpolation model from \citet{1995AJ....109.1044F}, which is marked by the solid line for each field.}
\label{fig:complete}
\end{figure*}

\begin{figure}[H]
  \begin{center}  
    \includegraphics{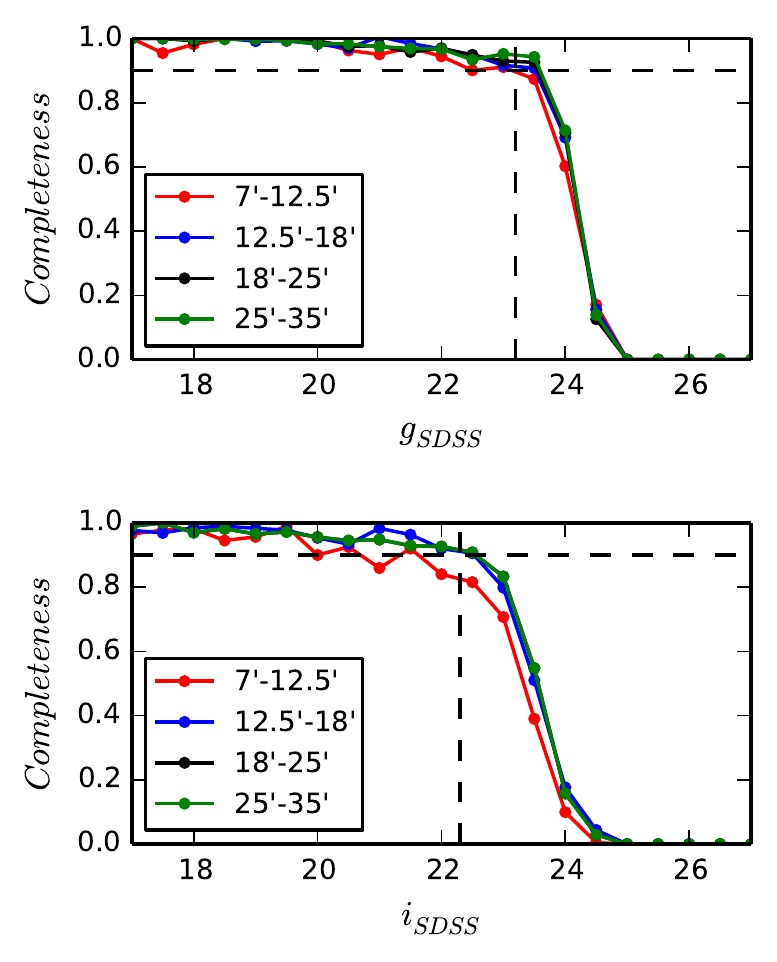}
  \end{center}
\caption{DECam completeness as a function of distance: $\textit{g}$-band completeness is displayed in the top panel and $\textit{i}$-band in the bottom panel. The vertical and horizontal dashed lines indicate the 90\% completeness level and the corresponding magnitude adopted. }
\label{fig:cdist}
\end{figure}

\subsection{Complete Catalogue}
Application of the completeness limits was the last step in obtaining our final photometric catalogues.  The resulting colour-magnitude diagrams (CMDs) are displayed in Figure \ref{fig:CMD}.  In both plots the main sequence and the main sequence turn-off (MSTO) of M2 are clearly seen, and it is these features that we focus on for the remainder of this work because they are the locations on the CMD where the signal of M2 populations is greatest with respect to background contamination. 

We performed photometric cuts to remove surplus stars in regions of the CMD that were not important for this study. Specifically, we removed the region occupied by red dwarfs in the foreground (belonging to the Galactic disk), which have $(g-i)>1.6$, as well as stars with an $\textit{i}$ magnitude brighter than $18$. This latter cut excluded the lower red giant branch of the cluster, but in this region of the CMD the number of M2 members relative to contaminants is low, especially at large radii from the cluster (this can be seen towards the top of the DECam CMD in Figure \ref{fig:CMD}). Also excluded are blue horizontal branch stars belonging to M2 -- although these are often used as tracers due to the low levels of contamination at blue colours, they are sufficiently bright that the majority of this population was saturated in all images such that the photometry was unreliable.

Finally, we note that the cluster centre, in both sets of imaging, is too crowded for us to retrieve any meaningful photometry. The effects of crowding can be observed by constructing the completeness function at different radii from the cluster centre.  For example, Fig. \ref{fig:cdist} displays the completeness function for the DECam data at different radii. Outside the nominal tidal radius of $\sim12.5^{\prime}$ arcmin there is no evident variation in the completeness function. Inside $12.5^{\prime}$, the completeness is noticeably degraded when we begin to include data at radii down to $\sim 7^{\prime}$, although note that above the $90\%$ cut-off that we assume across all pointings, the difference is marginal. By observing the radial dependence in this way, we set an inner limit of $7^{\prime}$ for the DECam data and $5^{\prime}$ for the MegaCam data. This provides an acceptable balance between probing more central regions of the cluster (for example, to accurately determine the locus of cluster populations on the CMD) and limiting the effects of crowding on the photometric uncertainties  and detection completeness. We emphasise, however, that our analysis is almost completely focused on regions well beyond the nominal tidal radius of $12.5^{\prime}$, where the spatial variation of the completeness curve is negligible.

\begin{figure*}
  \begin{center}  
    \includegraphics{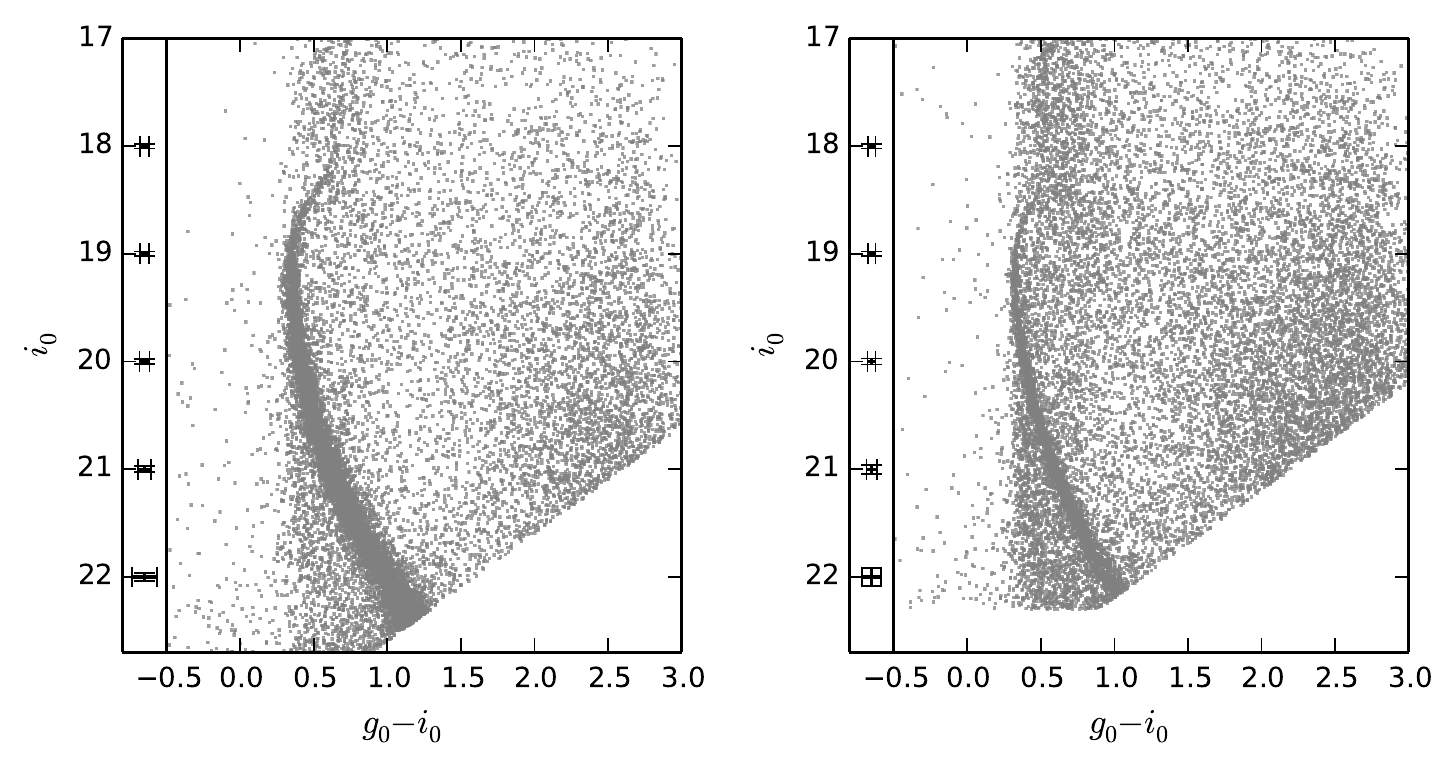}
  \end{center}
\caption{Colour-magnitude diagrams of our final stellar catalogues for MegaCam (left) and DECam (right). Both plots are accompanied by the typical photometric uncertainties at different brightness levels. Empty regions are caused by the $90\%$ completeness cuts that we applied. Note that for the DECam catalogue we only plot stars within $40\arcmin$ of the cluster centre to maintain visibility of the cluster sequences against the background.}
\label{fig:CMD}
\end{figure*}

\section{Results}
In this Section we describe our search for low surface brightness features in the vicinity of M2. Unless stated otherwise, the techniques we apply are identical for both the MegaCam and DECam photometric catalogues.

\subsection{Over-density Detection}

\subsubsection{Selection of Cluster Members}
The locus of M2 members is easily visible in both the MegaCam and DECam CMDs, and our aim is to reliably separate those M2 stars from the non-members -- primarily foreground stars belonging to the Milky Way. To do this, we adopted an isochrone from the Dartmouth Stellar Evolution Database\footnote{\url{http://stellar.dartmouth.edu/~models/index.html}} \citep{2008ApJS..178...89D} and fit it to the M2 sequence. We found that an isochrone with age$=13$ Gyr, $[Fe/H]=-1.7$, and $[\alpha$/$Fe] = +0.4$ provided a good description of the data  -- these parameters are a reasonable match for those in the literature \citep[e.g.,][]{2010ApJ...708..698D}. We adopted the absolute distance modulus listed in the 2010 edition of the \cite{1996AJ....112.1487H} catalogue but allowed small changes to obtain the best fit between the isochrone and the cluster main sequence. All stars in our catalogues were then assigned a "weight", according to a Gaussian distribution with the standard deviation set to be the colour difference from the isochrone value at a given $\textit{i}$ magnitude in units of the mean photometric uncertainty (determined from the rms of the photometry of stars observed in multiple images) in the measured colour at that magnitude. The Gaussian function was normalised such that a star falling on the isochrone would have a weight of $1.0$. Stars were then separated into two sets, "cluster" and "foreground", based on their assigned weight. The threshold used to separate stars into the two sets was determined empirically to encompass the observed width of the M2 main sequence, and corresponds to a weight value of $0.1$ for MegaCam and $0.2$ for DECam. Above these values stars are classified as belonging to the cluster; below them, to the foreground. Figure \ref{fig:contour} shows the results of our weighting scheme. Note that our set of cluster members still has some level of contamination due to non-members that happen, by chance, to lie near the isochrone. We attempt to account for this contamination in our subsequent analysis.

\begin{figure*}
  \begin{center}  
    \includegraphics{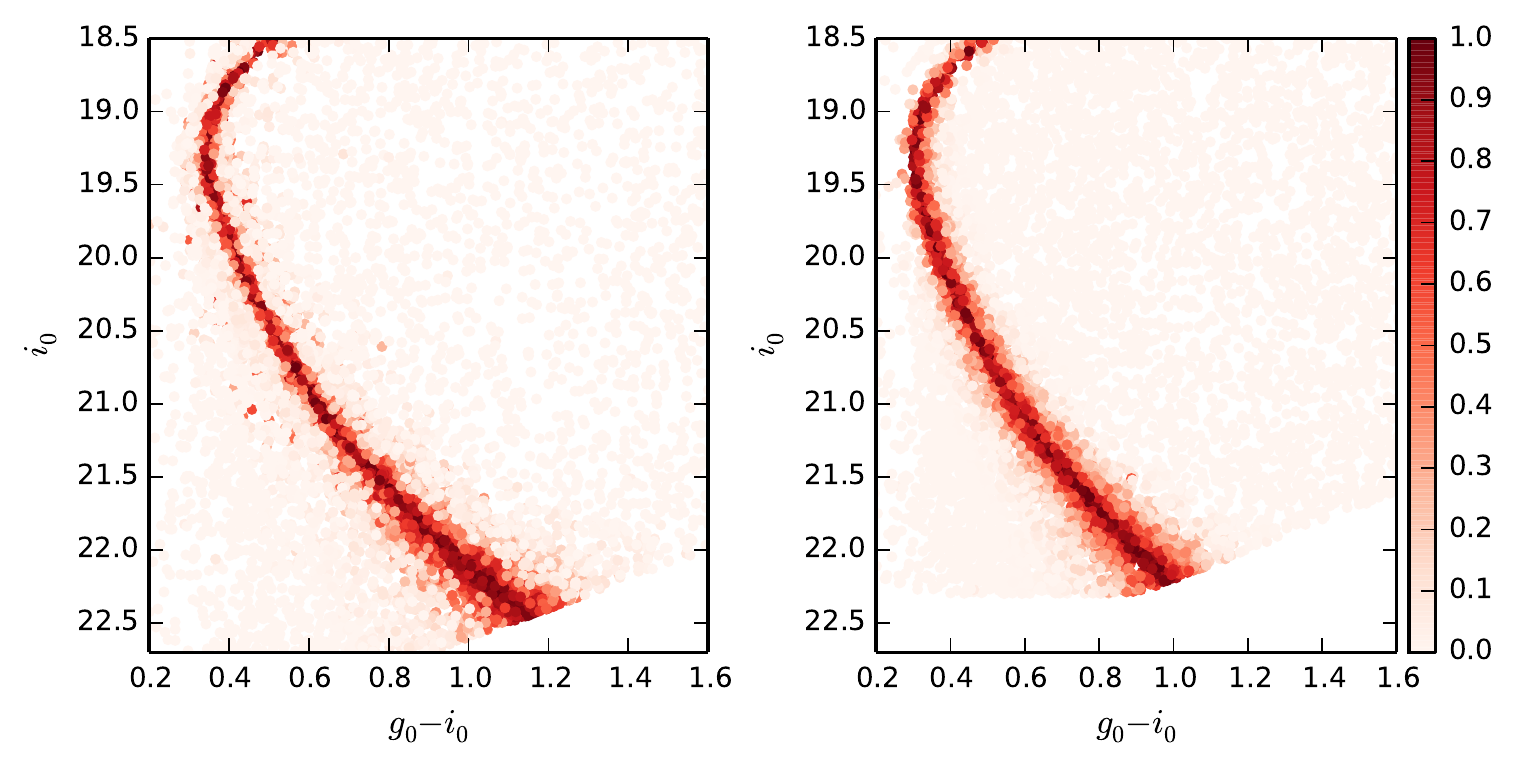}
  \end{center}
\caption{The isochrone-based weighting scheme for the CMDs show in Fig \ref{fig:CMD} -- each star has been coloured according to their assigned weight. As before, the CMD from our MegaCam catalogue is on the left, and from our DECam catalogue on the right.}
\label{fig:contour}
\end{figure*}

\begin{figure*}
  \begin{center}  
    \includegraphics{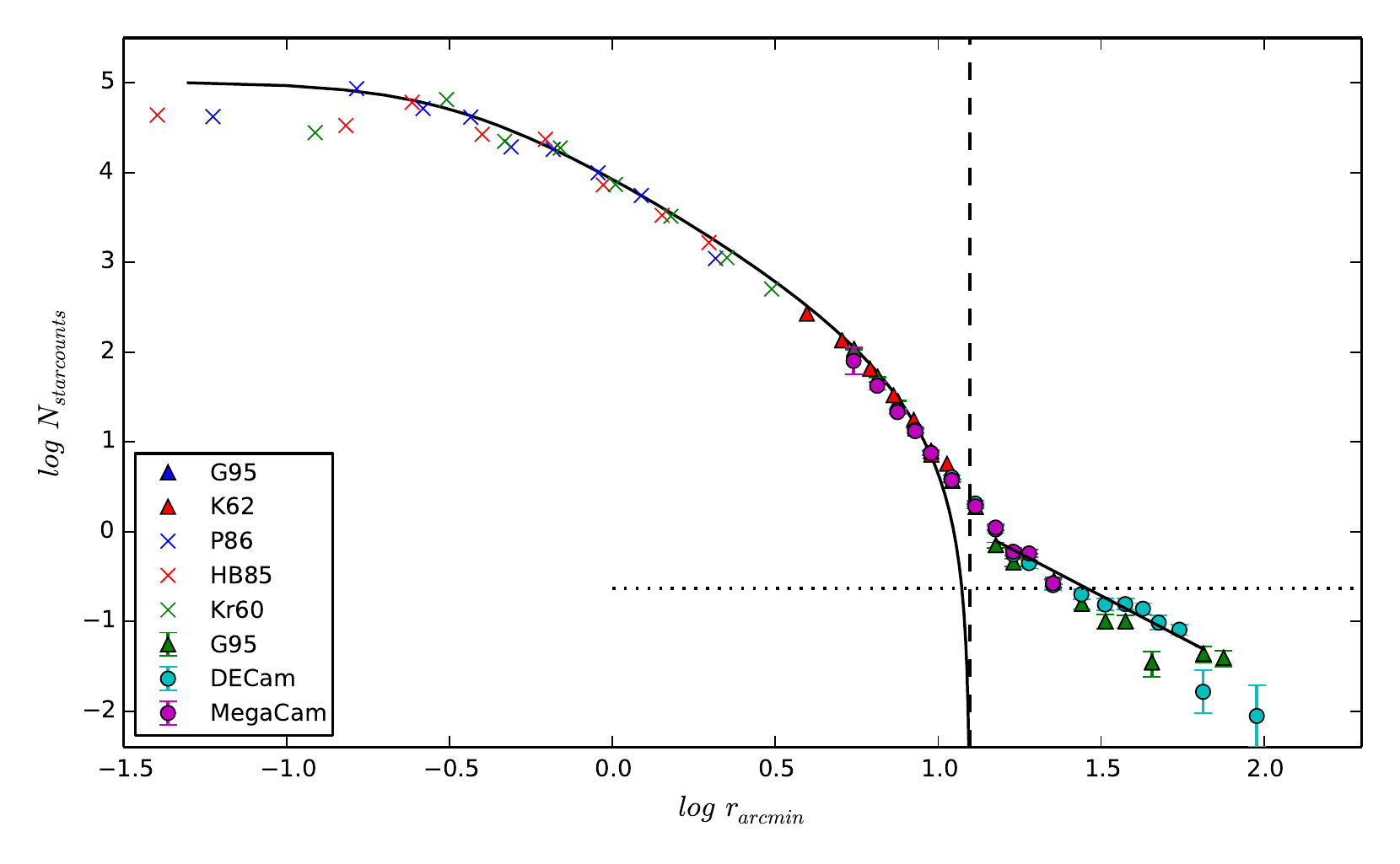}
  \end{center}
\caption{Our azimuthally averaged radial density profile for M2. This shows our measurements from both MegaCam and DECam, together with data from the literature. In general, solid symbols correspond to star count measurements, while crosses indicate aperture photometry. A \citet{1962AJ.....67..471K} profile, using structural parameters from \citet{1996AJ....112.1487H} (2010 edition), is marked with a solid black line. The nominal tidal radius of $12.5\arcmin$ for this model is indicated with a dashed vertical line, while the horizontal dotted line shows our calculated background level. Measurements in the cluster outskirts follow a power-law decline beyond the nominal tidal radius; that marked has an index of $\gamma = -2.2$. All literature measurements have been normalised to our MegaCam profile. \textit{Key: G95: \citep{1995AJ....109.2553G}, K62: \citep{1962AJ.....67..471K}, P86:\citep{1986PASP...98..192P}, HB85: \citep{1985MNRAS.214..491H}, Kr60: \citep{1960AJ.....65..581K}}.}
\label{fig:radprof}
\end{figure*}

\subsubsection{Radial Density Profile}\label{sec:RadProf}
Milky Way gobular clusters typically have radial density profiles that are well described by the family of (empirical) \citet{1962AJ.....67..471K} models:
\begin{equation}
n(r) = k\left(\frac{1}{\sqrt{1+\left(r/ r_c\right)^{2}}}-\frac{1}{\sqrt{1+\left(r_t / r_c\right)^{2}}}\right)^{2} \,,
\end{equation}
where $r_c$ and $r_t$ are the core and tidal radii respectively, and $r$ is the distance from the cluster centre. The coefficient $k$ is proportional to the central surface density (but is not the central density itself, as can easily be seen by setting $r$ to zero in the above equation). These models exhibit a characteristic sharp truncation as $r$ approaches $r_t$; \citet{1966AJ.....71...64K} later showed that such a truncation arises due to the influence of an external tidal field (in which case the velocity distribution takes a lowered Maxwellian form).  The signature of low surface brightness structure surrounding a globular cluster is the lack of this truncation; in such cases the outer density profile commonly exhibits a power-law decline -- for example due to tidal tails \citep[as observed around Pal 5][]{2001ApJ...548L.165O} or a diffuse envelope \citep[as seen around NGC 1851][]{2009AJ....138.1570O}. Stars in this region are commonly referred to as ``extra-tidal''.

To create a radial profile for M2, we split our catalogues into circular annuli about the cluster center, each of which was then sub-divided into eight sections. We calculated the density of cluster stars in each of these sub-sections and used the mean value as the annular density, and the standard deviation as the corresponding uncertainty in this value. We allowed the width of our annuli to increase with radius, to help suppress uncertainties due to the declining number of cluster stars at large distances from the cluster centre.  Also at large radii, portions of each annulus began to fall off the edge of our imaged mosaic, decreasing the effective area observed. To remedy this problem, for each impacted annular sub-section we performed a Monte Carlo simulation whereby a large number of points was uniformly generated in the region, and each point was determined to lie either inside, or outside, the field of view. We used the ratio of points that fell within the field to the total number placed to scale the calculated density to the correct level. If the ratio was less than $30\%$, it was considered to be too low and the corresponding ring section was disregarded from further analysis. 

We first created a profile without accounting for any residual contamination due to non-members of the cluster, and observed that the profile flattened to an approximately uniform value at a radial distance beyond $\approx 60\arcmin$. We estimated the foreground level by randomly sampling multiple sub-regions surrounding the cluster, $10\arcmin$ in diameter, centred at radii between $60\arcmin - 110\arcmin$, and generating a distribution of foreground densities. The foreground density ultimately subtracted from the profile was the mean of this distribution, and the uncertainty in this level was its standard deviation.

Our final radial density profile for M2 is plotted in Figure \ref{fig:radprof}. Our DECam measurements have been scaled to match those from MegaCam (which are deeper) by applying a vertical shift calculated in the region of overlap near the tidal radius. As our star counts do not sample the centre of the cluster, and we are unable to make integrated light measurements because the unresolved cluster centre is severely saturated in our images, we supplement our data with aperture photometry from \citet{1960AJ.....65..581K}, \citet{1985MNRAS.214..491H} and \citet{1986PASP...98..192P}. Since these were observations were made in different filters, we have applied a vertical shift to match them to our MegaCam star counts. Also plotted are star counts from \citet{1962AJ.....67..471K} and \citet{1995AJ....109.2553G}, again shifted to match our measurements. We further present a \citet{1962AJ.....67..471K} model, again normalised to our MegaCam data, using the core and tidal radii from \citet{1996AJ....112.1487H} (2010 edition): $r_c = 0.32\arcmin$ and $r_t = 12.5\arcmin$. 

It is immediately obvious from Figure \ref{fig:radprof} that our profile does not exhibit a sharp truncation, but instead follows a much more gradual decline with radius. Measurements by \citet{1995AJ....109.2553G} first presented possible evidence for extra-tidal features around M2, and this is strongly confirmed by our much higher quality data. According to our measurements the entire field of view of our MegaCam mosaic is occupied by M2 stars, even though the nominal tidal radius sits well within its footprint. Beyond the MegaCam observations, our DECam profile follows the findings of \citet{1995AJ....109.2553G} quite closely. The outer profile of M2 is reasonably well described by a power-law decline, with an index $\gamma = -2.2 \pm 0.2$.  In the next section we investigate how this extra-tidal structure is distributed on the sky.

\subsubsection{Foreground Subtraction and 2D Density Distribution}\label{sec:2DDist}
To explore the spatial density distribution of M2 members we created, for each camera, a 2D histogram of star counts using stars classified as cluster members according to their CMD weight. The number of bins along the spatial dimensions of the histogram (in this case $\alpha$ and $\delta$) is different between the two cameras, reflecting the different regions of the cluster the two different mosaics were focused on.  Table \ref{tab:param_detect} displays the bin sizes used for the two separate data sets.
  
As described above, each catalogue of cluster members still suffers from some degree of contamination. To account for this, we constructed a second 2D histogram of star counts for each camera, using stars classed as foreground members. For a given camera, both the ``cluster'' and ``foreground'' 2D distributions were normalized by dividing the number of stars in a bin by the total number of stars in the sample, then dividing by the area of the bin. We then fit a $1\times1$ bi-variate polynomial to the foreground distribution, and subtracted this from the density distribution of cluster stars to create a contamination corrected 2D density distribution, which had any large-scale gradients or fluctuations due to the foreground removed. The resulting maps were smoothed using a gaussian kernel of different widths for the two cameras.  The width of the smoothing function for both data sets is presented in Table \ref{tab:param_detect}.

Next, we searched these corrected distributions for regions harbouring over-densities of M2 stars. Our basic methodology was to define a region far from the cluster centre (as listed in Table \ref{tab:param_detect}), measure the mean and standard deviation of the bin densities in this region, and then examine fluctuations across the entire field of view in units of the number of standard deviations above or below the mean. For our DECam data this procedure was straightforward.  Based on our radial density profile we masked out everything within a radius of $60\arcmin$ of the cluster, and used all bins outside this radius to calculate the relevant statistics. We then experimented to determine what constituted a suitable threshold above the mean to consider a fluctuation as a {\it bona fide} over-density of M2 stars. If the threshold was set too low, too many peaks corresponding to random noise in the background were detected. Conversely, if the threshold was set too high, only the central region of the cluster was detected. Ultimately we explored a series of detection thresholds, beginning at 1$\sigma$ and extended to 3$\sigma$, to see how structures detected at lower significance were related to statistically more robust features.

For MegaCam, the situation was more complex. We previously observed that the entire field of view of our MegaCam mosaic is occupied by M2 stars. Hence, even the outskirts of the footprint did not constitute a clean non-cluster region for the purposes of determining the background statistics. As a result, while we went ahead and employed the same methodology using the region outside the nominal tidal radius of $12.5\arcmin$, we did not enforce a specific detection threshold when analysing the MegaCam results. Nonetheless, this allowed us to determine the overall shape of the distribution of M2 stars in the MegaCam footprint, and search for any substantial over-densities.

\begin{figure}
  \begin{center}  
    \includegraphics[width=\columnwidth]{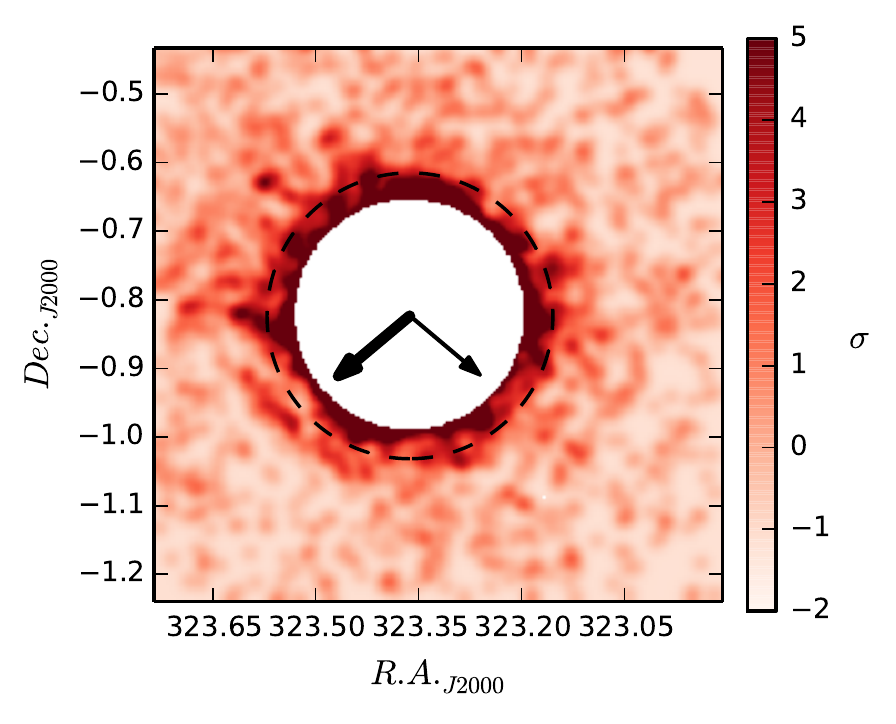}
  \end{center}
\caption{Stellar density distribution for the MegaCam catalogue, split into $7\arcsec \times 7\arcsec$ bins and smoothed using a gaussian kernel of width $35\arcsec$. The colour scale represents the number of standard deviations above the mean background value that a given bin sits. To enhance clarity in this map, a circular region of radius $10\arcmin$ has been masked at the cluster centre. The two arrows indicate the direction of the proper motion of M2 (the bold arrow) and the direction of the Galactic centre. The dashed ring indicates the nominal tidal radius of $12.5\arcmin$.}
\label{fig:MC_2d}
\end{figure}

\begin{figure*}
  \begin{center}  
    \includegraphics{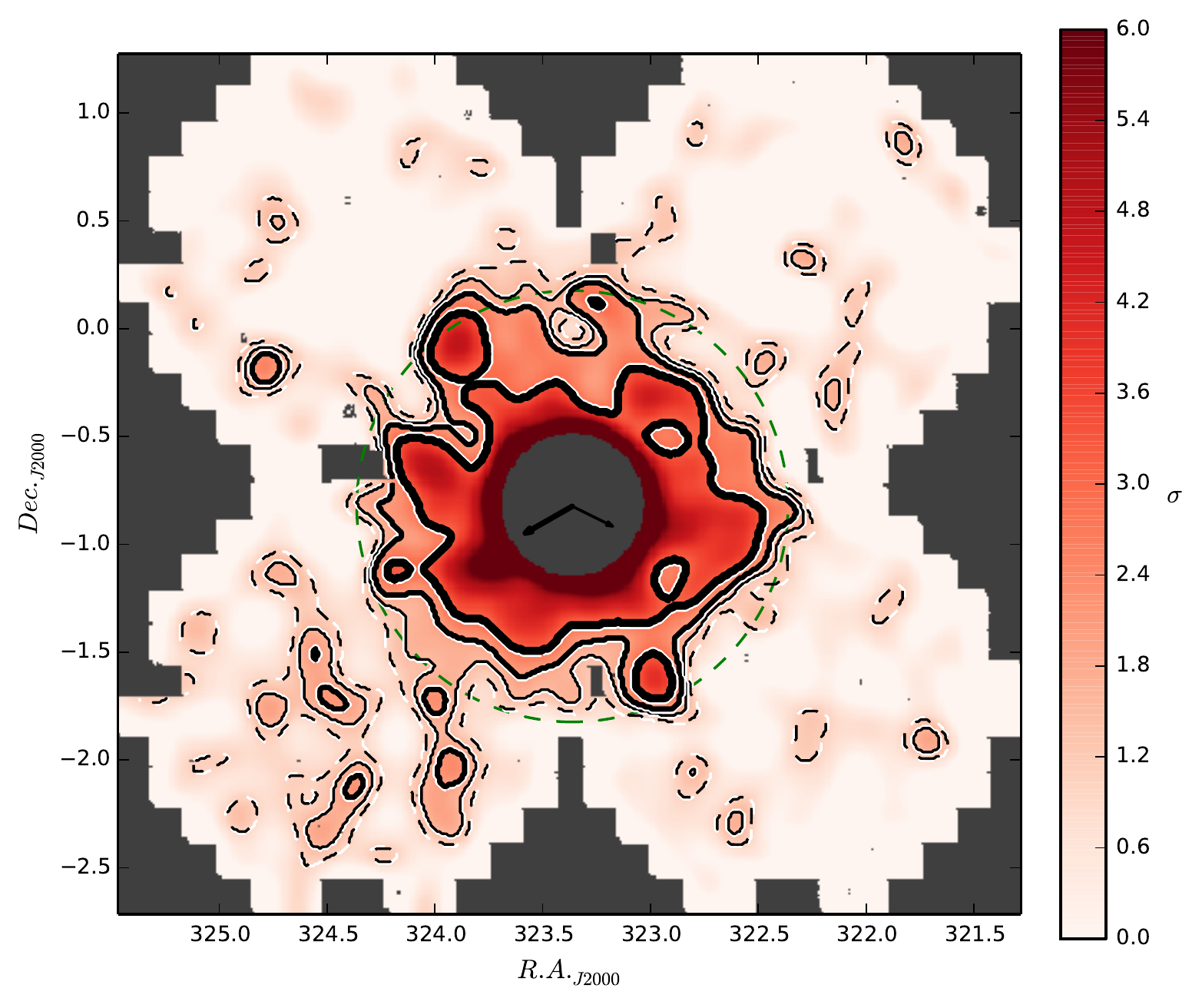}
  \end{center}
\caption{Stellar density distribution for the DECam catalogue split into $36\arcsec \times 36\arcsec$ bins and smoothed using a gaussian kernel of width $4.8\arcmin$. As for Figure \ref{fig:MC_2d} the colour scale represents the number of standard deviations above the mean background value that a given bin sits. The dashed contours indicate a level corresponding to $1\sigma$ above the mean bin density. Contours representing the $1.5\sigma$, $2\sigma$ and $3\sigma$ levels are shown by solid lines, increasing in thickness. A circular region of radius $20\arcmin$, almost twice the size of the nominal tidal radius, has been masked at the cluster centre. The outer dashed ring indicates a radius of $60\arcmin$. The arrows are the same as Figure \ref{fig:MC_2d}.}
\label{fig:DC_2d}
\end{figure*}

\begin{table}
\begin{center}
\caption{Parameters used to calculate the 2D density maps.}
\label{tab:param_detect}
\begin{tabular}{@{}ccc}
\hline \hline
Parameter&MegaCam &DECam\\
\hline
Bin width& $7\arcsec \times 7\arcsec$ & $36\arcsec \times 36\arcsec$\\
Smoothing width&$35\arcsec$&$4.8\arcmin$ \\
Masked region (radii) &$12.5\arcmin$ & $60\arcmin$\\
\hline
\end{tabular}
\end{center}
\end{table}

Our 2D density distribution maps are displayed in Figures \ref{fig:MC_2d} (MegaCam) and \ref{fig:DC_2d} (DECam). The first of these confirms that that stars belonging M2 can be observed across the entire area covered by our MegaCam imaging. Despite the difficulty in identifying a suitable region for determining the background statistics, the MegaCam map further reveals that the M2 stars are evenly spread, with no evident divergence from an approximately circular distribution, and no large scale over-densities. Moving to the DECam map, the extent of the envelope seen in the radial profile and the MegaCam map is revealed. We observe a large extended outer envelope, rather evenly spread in azimuth instead of constituting distinct tidal tails as suggested by \citet{1995AJ....109.2553G}. To the south-west, the envelope connects to a 3$\sigma$ detection through a low-significance feature, and consequently we present that over-density as part of the overall structure that we have detected. We find that the envelope extends to a radial distance of $\approx 60\arcmin$ ($\sim 210$ pc) at the $3\sigma$ threshold. While the radial profile hints at features possibly extending as far as $\sim 100\arcmin$ ($\sim 335$ pc), the overall shape of that potential structure is not evident from our 2D density distribution as it occurs at low significance. 

We employed a bivariate gaussian fit to the debris over the region $12.5\arcmin - 70\arcmin$ to explore the shape of the extended of M2 envelope and whether the structure has a distinct major axis direction. We performed this calculation using the python AstroML module\footnote{\url{http://www.astroml.org/}}, finding an ellipticity of $e = 0.11 \pm 0.06$ with the major axis oriented with a position angle $\theta = 69\degr  \pm 16\degr$ east of north. This ellipticity is a reasonable match for that determined for more central regions of the cluster, within $12.5\arcmin$, for which we find $e =  0.07 \pm 0.03$. However, the position angle of the major axis for this central region, $\theta = 138\degr  \pm 11\degr$ east of north, is somewhat different than for the envelope and may possibly indicate isophotal twisting (although this is not clearly evident from the density maps).

\begin{figure}
  \begin{center}  
    \includegraphics[width=\columnwidth]{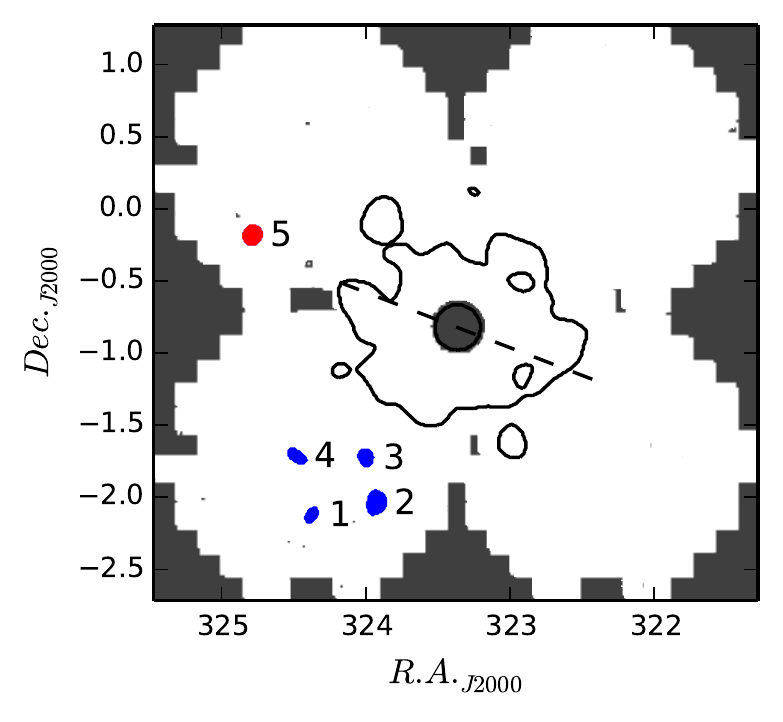}
  \end{center}
\caption{Over-density detections at the 2$\sigma$ level in the DECam map. The red detection is considered significant ($\zeta >3$), while the blue indicates $\zeta <3$ detections. The dashed line indicates the position angle of $69^{\circ}$. The envelope of M2 at the 3$\sigma$ level is also plotted. The binning and smoothing parameters are the same as for Figure \ref{fig:DC_2d}.   }
\label{fig:DC_OD_prior}
\end{figure}

\subsubsection{Significance of Individual Substructures}
Beyond the main envelope of M2 exists a number of over-densities detected at the $2\sigma$ level in the DECam map. We label, and show the locations of these regions, in Figure \ref{fig:DC_OD_prior}. Following \cite{2015ApJ...804..134R}, we employed a Monte Carlo simulation to investigate the significance of the number of cluster stars within any given over-density with respect to the typical number of stars obtained in a random sampling of the stellar catalogue. This allowed us to determine whether the number of cluster stars within the over-density was most likely just due to a random fluctuation in the field, or represented a grouping potentially related to M2.  To begin, the complete sample of DECam stars (cluster and foreground stars alike) was sorted into different sets based on their location inside an identified over-density or not. We defined a control group to be the set of all stars not located inside the 2$\sigma$ over-density in question, creating two sets of stars per over-density. For each region, we counted both the total number of stars and the number of these with weight $> 0.2$. Next, for a given over-dense region, we randomly selected the same total number of stars from the corresponding control sample, and determined the number of these with weight $> 0.2$. We repeated this sampling process $1000$ times per region, and then compared the observed number of stars with weight $> 0.2$ for a given region to the distribution bootstrapped from the control set. Specifically, we assigned each over-dense region a value, $\zeta$, defining how many standard deviations the true observed number of high-weight stars ($N_{OD}$) sits away from the mean of the control distribution ($\overline{N}_{CS}$)  \citep[see][]{2015ApJ...804..134R}:
\begin{equation}
\zeta = \frac{N_{OD}-\overline{N}_{CS}}{\sigma_{CS}}
\end{equation}
The mean and standard deviation ($\sigma_{CS}$) of each control distribution was determined by fitting a gaussian function to the sampled counts. The results of this significance testing are given in Table \ref{tab:statistics}. We deemed a detection to be significant if the number of high-weight stars in that over-density was $3\sigma$ or more above the mean of the control distribution (i.e. $\zeta > 3$); Only one of the five potential over-densities (number 5) was found to be significant. The CMD for the over-dense region is presented in Figure \ref{fig:M2_CMDd_sec}, together with the respective $\zeta$ value.

\begin{table}
\begin{center}
\caption{Over-dense regions and the results of our significance testing procedure.}
\label{tab:statistics}
\begin{tabular}{@{}ccccc}
\hline \hline
Detection & $N_{OD}$ & $\overline{N}_{CS}$&$\sigma_{CS}$&$\zeta$ \\
\hline
1&8&3.99&1.81&2.19\\
2&12&8.23&2.60&1.45\\
3&11&5.70&2.26&2.35\\
4&8&5.46&2.21&1.15\\
5&18&7.44&2.65&3.94\\
\hline
\end{tabular}
\end{center}
\end{table}

\section{Discussion}
\subsection{Nature of the Substructure Around M2}
Using deep imaging from MegaCam and DECam we have revealed the existence of an extended, diffuse stellar envelope surrounding the globular cluster M2. This structure extends to a radius of at least $60\arcmin$, or $\approx 210$ pc, from the centre of the cluster (according to the $3 \sigma$ contour in our DECam density map), and possibly as far as $\sim 100\arcmin$, or $\approx 335$ pc (according to our radial surface density profile). This corresponds to at least five times the nominal tidal radius of M2 from the literature \citep[see][]{1996AJ....112.1487H}. We find the envelope to be rather smooth and nearly circular -- its ellipticity is very mild ($e \approx 0.11$) and there is no obvious two-arm structure that might indicate the presence of classical tidal tails as seen around, for example, Palomar 5 or NGC 5466 \citep[e.g.,][]{2001ApJ...548L.165O,2006ApJ...637L..29B,2006ApJ...639L..17G}. This differs from to the conclusions of \citet{1995AJ....109.2553G}, who found extra-tidal stars surrounding M2, but suggested this was likely in the form of tidal tails.

The surface density of the envelope surrounding M2 follows a power-law decline with radius, of index $\gamma = -2.2 \pm 0.2$. Integrating the radial density profile allows us to estimate the ratio of mass in the envelope to the total mass of the cluster$+$envelope system. Examining our density profile (Figure \ref{fig:radprof}) we see that at radii smaller than $\sim 10\arcmin$ the literature \citet{1962AJ.....67..471K} model provides a good parametrization of the data; our new star counts begin diverging from the model outside this radius. We thus integrated the King model out to $10\arcmin$, and beyond this our DECam profile out to the $3\sigma$ detection limit of the envelope at $60\arcmin$. We consider everything outside the nominal literature tidal radius of $12.5\arcmin$ to constitute the envelope. With this definition, and ignoring the effect of mass segregation towards the cluster centre, our calculations reveal that the envelope comprises $\sim 1.6\%$ of the total mass of the cluster$+$envelope system.

One other Milky Way globular cluster, NGC 1851, is known to possess a substantial extended envelope component similar to that which we have revealed around M2 \citep[e.g.,][]{2009AJ....138.1570O,2014MNRAS.442.3044M}. The size of the envelope belonging to NGC 1851 is $\approx 250$ pc in radius, very similar to what we have observed for the envelope surrounding M2. It is also seen to follow a power-law decline in surface density with radius, although the slope may be shallower than we have observed for M2, with index $\gamma = -1.24 \pm 0.66$, and it likely contains a smaller fraction ($\sim 0.1\%$) of the total mass of the system \citep{2009AJ....138.1570O}.

Beyond the apparent edge of the M2 envelope, we have discovered a statistically significant over-density of cluster-like stars (detection 5; see Figure \ref{fig:M2_CMDd_sec}). On the sky this over-density is located along the axis suggested by the orientation of the major axis of the envelope, which sits at a position angle $\theta = 69\degr  \pm 16\degr$ east of north (see Figure \ref{fig:DC_OD_prior}). This may suggest a preferred axis for the overall M2 system; interestingly, this axis is quite well aligned with the direction of the Galactic centre from M2 (see Figure \ref{fig:DC_2d}). It is not clear whether the over-densitiy that we have detected might constitute an individual piece of M2 or its envelope, perhaps stripped via tidal forces, or whether it could represent a density peak in an even more extended envelope component that falls below the faint surface-brightness detection limit of our observations (note that the apparent ``edge'' of the envelope as seen in Figure \ref{fig:DC_2d} is due to our imposing a $3\sigma$ cut-off to the contouring, rather than actually comprising a physical boundary to the system). 

\begin{figure}
  \begin{center} 
  \includegraphics{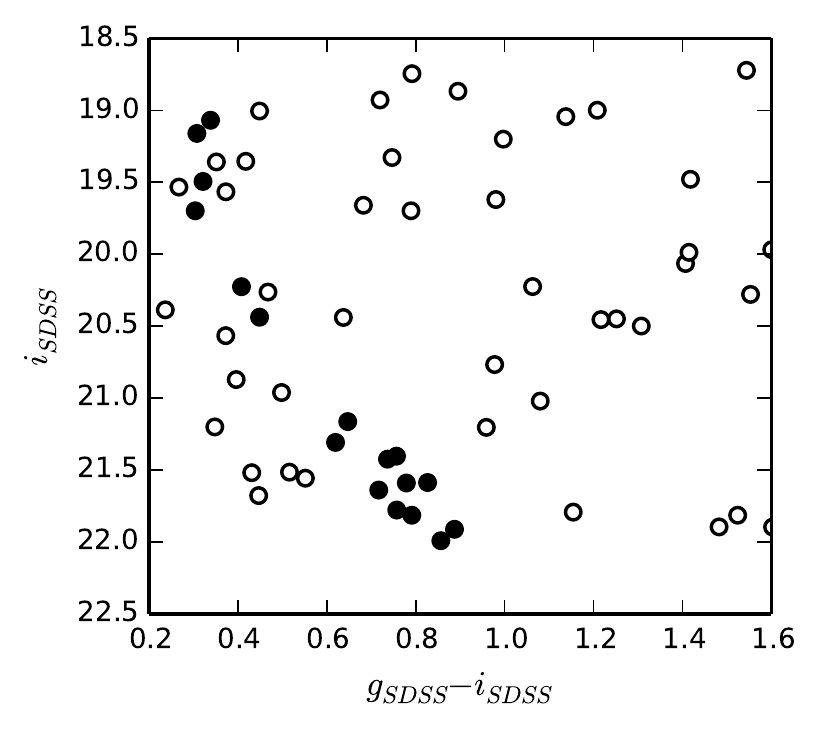}
   
  \end{center}
\caption{DECam CMD for detection 5 with a  $\zeta = 3.94$. The point style indicates the weight value -- filled points have weight $\geq 0.2$, while empty points correspond to weights $< 0.2$.}
\label{fig:M2_CMDd_sec}
\end{figure}

The origin of M2's diffuse stellar envelope is not clear from the presently available data. We can think of two simple scenarios: (i) the envelope is a natural product of the dynamical evolution of the cluster, perhaps driven by external tidal forces or shocks, or (ii) M2 is a globular cluster that belonged to, or was the nucleus of, a dwarf galaxy that was accreted by the Milky Way and destroyed, leaving behind the cluster$+$envelope system. 

We first explore the possibility that the envelope is a product of the dynamical evolution of M2. Proper motion measurements and orbital models from \citet{1999AJ....117.1792D} place M2 on a rather elliptical orbit ($e\approx 0.7$) with a period of $\sim650-850$ Myr, a perigalactic radius of $\sim 6$ kpc, and an apogalactic radius of up to $\sim 40$ kpc. Assuming this has not evolved significantly in the past, over a Hubble time M2 would have traversed of order $\approx 15$ orbits, and, as a consequence, has undergone multiple disk passages and shocks. Such events are known to accelerate the escape of stars from clusters, and hence speed up their dynamical evolution and ultimate disruption \citep[e.g.,][]{1997ApJ...474..223G}. 

Models of globular cluster evolution show that stars that become energetically unbound cross the Jacobi radius (where the internal gravitational acceleration equals the tidal acceleration) through the Lagrange points to form tidal tails which can be very long but have a width roughly equivalent to that of the cluster \citep[see e.g.,][and references therein]{1999A&A...352..149C, 2010MNRAS.407.2241K, 2011MNRAS.418..759R}. A number of striking examples are known in the Milky Way halo -- for example, Palomar 5 and NGC 5466 as noted above. M2 presently sits about $7$ kpc below the Galactic plane, and has a large velocity component in the negative $Z$ direction \citep[i.e., away from the plane][]{1999AJ....117.1792D}. Hence it is likely that M2 has recently passed through perigalacticon and suffered a disk shock, such that its extended envelope, and indeed the more remote over-densities, might plausibly reflect a wave of escaping stars. However, we find no evidence for narrow tidal tails, -- the envelope is rather evenly distributed in azimuth and is, in any case, {\it much} wider than the cluster.

Models of the formation of tidal tails \citep[see e.g.,][]{2010MNRAS.401..105K} show that stars with sufficient energy to escape the cluster can take many dynamical times to move through the Lagrange point. During this stage these stars preferentially populate the outermost regions of the cluster and can form a halo-like structure that deviates from a King profile around the Jacobi radius. However, we do not believe the envelope of M2 is due to this type of process. \cite{2010MNRAS.401..105K} find that except for clusters near core collapse, the King tidal radius fitted from a surface density profile is in general a reasonably close approximation to the Jacobi radius, with $r_t/r_J$ in the range $\sim 0.8-1.2$. For M2, we find that the tidal radius of $12.5\arcmin$ listed in the Harris catalogue provides a good description of the surface density profile; indeed we observe deviation to a power-law profile to begin at approximately this radius. \cite{2010MNRAS.401..105K} further show that beyond the Jacobi radius, unbound material tends to obey a power-law fall-off with a slope of $\gamma \sim -4$ to $-5$. Steep profiles like this are seen for many globular clusters \citep[e.g.,][]{2012MNRAS.419...14C}, but we observe a much shallower profile with $\gamma = -2.2$ for M2. \cite{2010MNRAS.401..105K} find that clusters near apogalacticon can have shallower power-law indices up to $\gamma \sim -1$. Note however that, according to the orbit calculation by \citet{1999AJ....117.1792D} and \citet{2006ApJ...652.1150A}, M2 should be currently far from apogalacticon, although significant uncertainties in its actual orbital path are present.

Simulations modelling the formation of tidal tails \citep[e.g.,][]{2006MNRAS.367..646L} show that the debris lost from a cluster ought to appear spatially elongated at a few Jacobi radii from the centre. However, we do not observe substantial elongation of the M2 envelope out to $\sim 5$ times the Jacobi radius. While we cannot rule out the possibility that we are seeing tidal tails lying along, or close to, the line-of-sight vector, i.e., seen end on, this projection is statistically unlikely.

It is also relevant that that M2 is not particularly vulnerable to disk shocks due to its relatively high mass; \citet{1997ApJ...474..223G} find that the combined effect of disk and bulge shocks on M2 (as quantified by the ``destruction rate'' due to these processes) is comparable to that of internal two-body relaxation \citep[see also][]{1999AJ....117.1792D, 2006ApJ...652.1150A}. 

Given that M2 spends a large proportion of its orbit at much larger Galactocentric radii than where it is presently located, it is reasonable to ask whether evolution in a more benign environment might facilitate the production of a diffuse envelope. It is known that very isolated clusters tend to build up a surrounding halo of stars that have been scattered on to radial orbits by two- or three-body encounters in the inner regions of the cluster, and that the resulting density profile ought to possess a power-law decline, in projection, of index $-2.5 \la \gamma \la -2.3$ \citep[see e.g., the discussion in][]{2010MNRAS.401..533M}. This is quite similar to what we observe for the envelope of M2; moreover, the time-averaged tidal radius for the cluster would be a factor of several larger than at its present location, which might allow the envelope to become populated. Arguing against this scenario is that it takes many relaxation times to establish the core-halo structure, and, furthermore, it is not clear whether this would survive repeated pericentre passages and disk shocks. It is relevant that the half-mass relaxation time for M2 is $\sim 2.5$ Gyr \citep{1996AJ....112.1487H}, which is substantially longer than its orbital period.

We now turn to the possibility that M2 was once part of a dwarf galaxy that was accreted and destroyed by the Milky Way. This hypothesis has previously been advanced to explain the envelope surrounding NGC 1851 \citep{2009AJ....138.1570O}, and the abundance patterns observed for stars in the envelope of NGC 1851 are compatible with this idea \citep{2014MNRAS.442.3044M}. Simulations performed by \citet{2012MNRAS.419.2063B} have demonstrated that a diffuse envelope can indeed form around the compact nucleus of a dwarf galaxy after the original host has been largely stripped away by tidal forces. M2 shares a number of unusual attributes in common with other Milky Way globular clusters that have been suggested to be remnant dwarf nuclei. In particular, it exhibits an internal dispersion in iron abundance in the form of three distinct stellar populations \citep{2014MNRAS.441.3396Y}, that further subdivide into sub-populations according to variations in s-process element abundances, light element abundances, and helium abundances \citep{2012A&A...548A.107L,2013MNRAS.433.1941L,2015MNRAS.447..927M}. In this regard it is similar to $\omega$ Cen \citep[e.g.,][]{2014ApJ...791..107V}, which has long been hypothesized to be a former dwarf galaxy nucleus; to M54 \citep[e.g.,][]{2007ApJ...667L..57S,2010A&A...520A..95C}, which is either the nucleus, or central globular cluster, of the Sagittarius dwarf \citep{1995MNRAS.277..781I,2008AJ....136.1147B}; and indeed to NGC 1851 \citep[e.g.,][]{2010ApJ...722L...1C,2015MNRAS.446.3319Y}. It is also relevant that the overall size of the envelope that we have observed around M2, with a radius of {\it at least} $\sim 210$ pc, is not dissimilar in size to the half-light radii of many typical dwarf galaxies in the Local Group \citep{2012AJ....144....4M}.

The number of stripped dwarf nuclei with masses between $10^{5}-10^{6}$ M$_{\odot}$ in the Milky Way halo has been proposed by \citet{2014MNRAS.444.3670P} to be between one and three, based on the Millennium II simulation and semi-analytic modelling. However, as noted by these authors, this is lower than the number of objects already hypothesised to be stripped nuclei of this type. According to the criterion specified by \cite{2014MNRAS.444.3670P} -- that a globular cluster which is a former dwarf nucleus ought to have an internal spread in age and/or heavy element abundances -- and the discussion above, M2 should also be considered a member of this category, increasing the possible tension between simulation and observation.  However, the authors note that the Poisson uncertainties on their estimate are substantial, and could accommodate a larger number of systems.  Moreover, it is not clear that their specified criterion uniquely identifies stripped dwarf nuclei.  It is possible that the presence of an extended outer structure, as we have observed for M2, could constitute an additional marker.

If it is true that M2 was once a member of a now-defunct dwarf, the lack of a large stellar stream in the vicinity of the cluster (as is seen, for example, for the disrupting Sagittarius dwarf) may suggest that the dwarf galaxy that housed M2 was accreted very long ago. In this respect, the over-density that we have detected beyond the main envelope is potentially the only remaining fragment of that stellar stream in our field of view. As noted above, this over-density might also signify the presence of an even more extended envelope -- perhaps stream-like in nature -- that connects it to M2 but falls below the faint surface brightness threshold of our observations. In this regard, probing even further down the M2 main sequence could help detect such a feature, although the fact that this will have to be done over a relatively large area of sky might mean that we will need to wait for the advent of facilities such as the Large Synoptic Survey Telescope \citep[LSST,][]{2008arXiv0805.2366I}. Apart from this, spectroscopic follow-up of stars in the M2 envelope and the nearby over-density should help identify whether the envelope exhibits abundance patterns similar to those of the cluster, and confirm whether the over-density is truly related to the cluster or not.

\section{Conclusions}
We have searched the region surrounding the Milky Way globular cluster M2 for the presence of low surface brightness substructures, using deep wide-field imaging mosaics from MegaCam and DECam. We use the observed colour-magnitude diagram to identify likely cluster members across the respective fields of view, and find that a composite radial surface density profile indicates substantial extra-tidal populations extending well beyond the literature value for the tidal radius of $12.5\arcmin$. The surface density declines with radius according to a power law with index $\gamma =  -2.2 \pm 0.2$. These remote M2 populations entirely fill our $0.8\degr \times 0.8\degr$ MegaCam mosaic, and it is only with a $\sim 13$ square degree mosaic from DECam that we are able to identify a diffuse, extended envelope surrounding the cluster to a radial distance of at least $60\arcmin$ ($\sim 210$ pc),  five times larger than the nominal tidal radius. Our two-dimensional density map reveals the envelope to be mildly elliptical, with $e = 0.11 \pm 0.06$ and the major axis oriented at a position angle of $\theta = 69^{\circ}  \pm 16^{\circ}$ east of north. There is no evidence for a distinct stellar stream or tidal tails, although we identify a small but statistically significant over-density of M2 stars beyond the apparent edge of the envelope, that follows a potential axis extending from north-east to south-west in broad agreement with the orientation of the envelope.

The nature and origin of the diffuse envelope surrounding M2 is not well understood. One possibility is that this structure is due to the dynamical evolution of the cluster, although how external factors such as tidal shocking might give rise to such an envelope, as opposed to the distinct tidal tails observed around disrupting globular clusters and seen in numerical simulations, is not clear. Numerous globular clusters have been found with power-law extended profiles \citep[e.g.,][]{2011MNRAS.417.2411C,2012MNRAS.419...14C,2014MNRAS.445.2971C,2005ApJS..161..304M} without tidal tails, though none of these studies have a found an envelope to the size of, or exhibiting a profile a shallow as, M2. An alternative scenario is that M2 was originally formed in a dwarf galaxy that was later accreted into the Milky Way halo and destroyed -- in this case the envelope might constitute the final remaining vestiges of the host.  A similar structure has been observed to surround the globular cluster NGC 1851 \citep[e.g.,][]{2009AJ....138.1570O,2014MNRAS.442.3044M}, and simulations of this system have shown that the nucleus of a dwarf galaxy can possess a halo-like structure surrounding the dense core long after the majority of the original dwarf and its dark matter halo have been stripped away and lost \citep{2012MNRAS.419.2063B}. In this context it is intriguing that M2 is a member of a small group of massive Milky Way globular clusters (also including NGC 1851) observed to exhibit internal dispersions in both iron abundance and s-process elements \citep[e.g.,][]{2014MNRAS.441.3396Y}. 
Deeper imaging of the region around M2, together with spectroscopic velocity and abundance measurements of stars in the envelope, will be required to understand the origin of this structure with greater certainty.

\section{Acknowledgments}
PBK and TAR acknowledge financial support through Australian Postgraduate Awards. GDC and ADM are grateful for support from the Australian Research Council (Discovery Projects DP120101237 and DP150103294). We thank the referee for their suggestions which have improved the quality of this paper.

This paper includes data gathered with the 6.5 meter Magellan Telescopes located at Las Campanas Observatory, Chile. Australian access to the Magellan Telescopes was supported through the Collaborative Research Infrastructure Strategy of the Australian Federal Government.

 This project used data obtained with the Dark Energy Camera (DECam), which was constructed by the Dark Energy Survey (DES) collaboration. Funding for the DES Projects has been provided by 
the U.S. Department of Energy, 
the U.S. National Science Foundation, 
the Ministry of Science and Education of Spain, 
the Science and Technology Facilities Council of the United Kingdom, 
the Higher Education Funding Council for England, 
the National Center for Supercomputing Applications at the University of Illinois at Urbana-Champaign, 
the Kavli Institute of Cosmological Physics at the University of Chicago, 
the Center for Cosmology and Astro-Particle Physics at the Ohio State University, 
the Mitchell Institute for Fundamental Physics and Astronomy at Texas A\&M University, 
Financiadora de Estudos e Projetos, Funda{\c c}{\~a}o Carlos Chagas Filho de Amparo {\`a} Pesquisa do Estado do Rio de Janeiro, 
Conselho Nacional de Desenvolvimento Cient{\'i}fico e Tecnol{\'o}gico and the Minist{\'e}rio da Ci{\^e}ncia, Tecnologia e Inovac{\~a}o, 
the Deutsche Forschungsgemeinschaft, 
and the Collaborating Institutions in the Dark Energy Survey. 
The Collaborating Institutions are 
Argonne National Laboratory, 
the University of California at Santa Cruz, 
the University of Cambridge, 
Centro de Investigaciones En{\'e}rgeticas, Medioambientales y Tecnol{\'o}gicas-Madrid, 
the University of Chicago, 
University College London, 
the DES-Brazil Consortium, 
the University of Edinburgh, 
the Eidgen{\"o}ssische Technische Hoch\-schule (ETH) Z{\"u}rich, 
Fermi National Accelerator Laboratory, 
the University of Illinois at Urbana-Champaign, 
the Institut de Ci{\`e}ncies de l'Espai (IEEC/CSIC), 
the Institut de F{\'i}sica d'Altes Energies, 
Lawrence Berkeley National Laboratory, 
the Ludwig-Maximilians Universit{\"a}t M{\"u}nchen and the associated Excellence Cluster Universe, 
the University of Michigan, 
{the} National Optical Astronomy Observatory, 
the University of Nottingham, 
the Ohio State University, 
the University of Pennsylvania, 
the University of Portsmouth, 
SLAC National Accelerator Laboratory, 
Stanford University, 
the University of Sussex, 
and Texas A\&M University. 

Funding for the Sloan Digital Sky Survey (SDSS) has been provided by the Alfred P. Sloan Foundation, the Participating Institutions, the National Aeronautics and Space Administration, the National Science Foundation, the U.S. Department of Energy, the Japanese Monbukagakusho, and the Max Planck Society. The SDSS Web site is http://www.sdss.org/.

The SDSS is managed by the Astrophysical Research Consortium (ARC) for the Participating Institutions. The Participating Institutions are The University of Chicago, Fermilab, the Institute for Advanced Study, the Japan Participation Group, The Johns Hopkins University, Los Alamos National Laboratory, the Max-Planck-Institute for Astronomy (MPIA), the Max-Planck-Institute for Astrophysics (MPA), New Mexico State University, University of Pittsburgh, Princeton University, the United States Naval Observatory, and the University of Washington.

\bibliographystyle{mnras}
\bibliography{M2_clean}

\label{lastpage}
\end{document}